\documentclass[seceq,preprint]{ptptex}

\usepackage[dvips]{graphicx}

\newcommand{\ol}{\overline}
\newcommand{\wt}{\widetilde}

\def\Im{\mathop{\rm Im}}

\preprintnumber[3cm]{
UT-02-41\\{\tt hep-th/0208079}\\August, 2002}

\title{
Open String-BMN Operator Correspondence\\in the Weak Coupling Regime%
}

\author{
Yosuke \textsc{Imamura}\footnote{e-mail: imamura@hep-th.phys.s.u-tokyo.ac.jp}%
}

\inst{
Department of Physics,  Faculty of Science,\\
University of Tokyo, Tokyo 113-0033, Japan
}

\abst{
We reproduce the correspondence between open string states $|n\rangle$
and BMN operators $O_n$ in ${\cal N}=2$ supersymmetric gauge theory
using a worldsheet computation with the relation $\langle B|n\rangle=O_n$,
where $\langle B|$ is a boundary state of D3-branes.
We regard the NS ground states and their excitations
with respect to a bosonic oscillator
as the states $|n\rangle$ and obtain a chiral operator and
BMN operators, respectively,
in correspondence to them.
Because we take the coupling constant to be weak and the background spacetime
to be flat Minkowski, not AdS, the string states $|n\rangle$ are
circular waves around the D3-branes, instead of KK modes on ${\bf S}^5$.
We also discuss the $1/J$ correction to the correspondence.
}

\begin{document}

\maketitle

\section{Introduction}
String theory is a powerful tool for investigating the dynamics of gauge theories.
This is due to the many kinds of dualities that relate
string theory on various backgrounds and gauge theories.
One of the most important dualities recently attracting great interest is
AdS/CFT duality\cite{adscft}.
With this duality, we can compute perturbative and non-perturbative quantities in gauge theories
by using perturbative string or supergravity.
One important example is the correlation functions in gauge theory.
It is believed that there is a one to one correspondence between
gauge singlet operators in CFT and fields propagating in the bulk.
If we could determine this correspondence, we could compute correlation functions of the
gauge singlet operators by analyzing the propagation of the corresponding
fields in the bulk.\cite{GKP,holography}

In the case of AdS$_5$/${\cal N}=4$ SYM duality,
the correspondence between gauge singlet operators and string states in
the AdS background can, in principle, be determined
in the following way.
Let us consider strings on the D3-brane background.
In the region of large $r$, the background spacetime is approximately flat Minkowski.
Let $|n\rangle_{\rm circ}$ be a string state on this flat spacetime.
Because we consider eigenstates of angular momentum,
we use circular wave states, $|n\rangle_{\rm circ}$.
In the weak coupling regime, D3-branes can be treated as `branes' playing the role of boundary condition of strings,
and we can define a map from a circular wave state $|n\rangle_{\rm circ}$ to
a gauge singlet operator $O_n$ in the gauge theory by the relation
\begin{equation}
\langle B|n\rangle_{\rm circ}=O_n,
\label{circ-on}
\end{equation}
where $\langle B|$ is the boundary state representing the coinciding D3-branes.
We define this boundary state as a surface state.
Because $\langle B|$ carries the information of fields on the D3-branes,
the product $\langle B|n\rangle_{\rm circ}$ is a gauge singlet function of the fields on
D3-branes.

In contrast to the above case, in the strong coupling regime, the background geometry near horizon is
well-approximated as an anti-de Sitter spacetime.
Let $|n\rangle_{\rm AdS}$ denote a string state in the AdS background.
If we could solve the string wave equation on the entire D3-brane geometry,
we would obtain the correspondence
\begin{equation}
|n\rangle_{\rm circ}\quad\leftrightarrow\quad|n\rangle_{\rm AdS},
\label{circ-ads}
\end{equation}
by interpolating between the wave functions of $|n\rangle_{\rm circ}$ and $|n\rangle_{\rm AdS}$
with a wave function on the D3-brane geometry.
Combining the two relations (\ref{circ-on}) and (\ref{circ-ads}), we could obtain the correspondence
between string states in the AdS background and gauge singlet operators.

It was recently shown that the string spectrum
in a PP-wave background, which is obtained as the Penrose limit of the AdS background,
can be solved exactly in the light-cone formalism\cite{metsaev,MT}.
Berenstein, Maldacena and Nastase (BMN) suggested a correspondence between these string states and gauge singlet operators,
which is now reffered to as BMN operators\cite{BMN}.
They computed the anomalous dimensions of the single trace operators
possessing large R-charge and
showed that they coincide with the energy spectrum of closed strings
in the PP-wave background.
Since the work by BMN,
many papers have appeared studying the extension of this correspondence into orbifold\cite{enhanced,GO,ZS,takatera,ASJ,quiver,FK,moose,OhTatar}
and open strings\cite{chuho,open,LeePark}.
The interaction of strings in the PP-wave and its relation to Yang-Mills interaction are
studied in Refs.\citen{doublescaling,bn,constable,check,threept}.

The purpose of this paper is to study this correspondence in the manner explained above.
We determine the map between string circular wave states
in the flat background and gauge singlet operators as the relation (\ref{circ-on}).
For closed string and single trace operators, this was already done in Ref.\citen{imamura}.
In this paper, we consider the D3-D7 system and repeat the analysis given in \citen{imamura} for 7-7 open strings and
non-trace operators.

Unfortunately, we have not succeeded in constructing the string theory on
the AdS$_5$ background, to say nothing of the D3-brane background.
Therefore, we cannot obtain the exact correspondence between $|n\rangle_{\rm circ}$ and $|n\rangle_{\rm AdS}$.
However, if we assume that the typical scale of the background geometry is sufficiently large
compared to the typical scale of the string oscillations,
we can approximately separate the string oscillation from the motion of the center of mass.
This implies that string wave functions can be decomposed into two factors representing the motion of the center of mass
and oscillations, respectively.
Therefore, we can obtain $|n\rangle_{\rm AdS}$ corresponding to $|n\rangle_{\rm circ}$ by replacing
the wave function for the center of mass of the circular wave $|n\rangle_{\rm circ}$
by a wave function in the AdS background.
Combining this and the relation between $|n\rangle_{\rm circ}$ and the $O_n$ we obtain qubsequently using (\ref{circ-on}),
we obtain an operator-string state correspondence quite similar to that suggested in Ref.\citen{open}.

Because we need to use global coordinates in order to take the Penrose limit,
the gauge theory corresponding to the PP-wave is not a theory on ${\bf R}^4$.
This problem is discussed in several works
\cite{DGR,LOR,KirPio}.
In Ref.\citen{bn}, it is clearly explained
that the corresponding gauge theory is compactified on ${\bf S}^3$
and only the KK lowest modes appear in the PP-wave/gauge theory correspondence.
This implies that the PP-wave is dual to a certain matrix theory.\cite{DSR,ver,Soch,SugiYoshi1,okuyama,SugiYoshi2}
Because in this paper we discuss the duality between ${\cal N}=4$ theory on ${\bf R}^4$
and string states, the correspondence obtained here should not be directly identified with the PP-wave/gauge theory correspondence suggested by BMN.

This paper is organized as follows.
In \S\ref{gauge.sec}, we briefly review the computation of
anomalous dimensions of gauge singlet operators with $\Delta_{\rm tree}-J=1$ and $2$
and the correspondence between these operators and open string states.
In section \ref{string.sec},
we study the coupling of 7-7 open string circular wave states and D3-branes.
We find that this coupling precisely reproduces
the operator-string state correspondence
expected from the gauge theory analysis given in  \S2.
We also discuss the $1/J$ correction to the correspondence.
The last section is devoted to conclusions.

\section{${\cal N}=2$ SQCD}\label{gauge.sec}
\subsection{Fields and symmetries}
In this section, we define the gauge theory we will discuss in this paper
and briefly review how the BMN operators are determined by computing two-point functions.\cite{open,LeePark,pporientifolds}.

We consider ${\cal N}=2$ $U(N_c)$ Yang-Mills theory
with one adjoint and $N_f$ fundamental hypermultiplets
and $Sp(N_c/2)$ Yang-Mills theory obtained by ${\bf Z}_2$ projection
from the $U(N_c)$ theory.
[Of course, $N_c$ should be an even integer when we consider the $Sp(N_c/2)$ theory.]
The $U(N_c)$ theory is realized on $N_c$ coinciding D3-branes on a background with $N_f$ D7-branes.
The ${\bf Z}_2$ projection to obtain the $Sp(N_c/2)$ theory
is equivalent to the introduction of an orientifold $7$-plane parallel to the D7-branes.

We assume that the directions of the D3-branes and D7-branes are
0123 and 01234567, respectively,
and that they are located at the origin of the transverse coordinates.
We divide the ten coordinates $X^I$ ($I=0,\ldots,9$)
into three groups:
$X^\mu$ ($\mu=0,1,2,3$), $X^i$ ($i=4,5,6,7$)
and $W=(X^8+iX^9)/\sqrt2$.
We call these three types of directions
the NN-, DN- and DD-directions, respectively.
We also use complex coordinates composed of the DN-directions
$Z=(X^4+iX^5)/\sqrt2$ and $\wt Z=(X^6+iX^7)/\sqrt2$.
(Table \ref{branes.tbl})
\begin{table}[htb]
\caption{The directions of the D3- and D7-branes.
The angular momenta $J$, $\wt J$ and $J_W$ are defined as
charges associated with rotations in $45$, $67$ and $89$ planes,
respectively.}
\label{branes.tbl}
\begin{center}
\begin{tabular}{c|cccc|cc|cc|cc}
\hline
\hline
       & 0 & 1 & 2 & 3 & 4 & 5 & 6 & 7 & 8 & 9 \\
\hline
       &\multicolumn{4}{c|}{$X^\mu$} & \multicolumn{2}{c|}{$Z$}
       & \multicolumn{2}{c|}{$\wt Z$} & \multicolumn{2}{c}{$W$} \\
\hline
D3     & $\circ$ & $\circ$ & $\circ$ & $\circ$ &&&&&&\\
D7, O7 & $\circ$ & $\circ$ & $\circ$ & $\circ$
       & $\circ$ & $\circ$ & $\circ$ & $\circ$ \\
\hline
\end{tabular}
\end{center}
\end{table}

The $U(N_c)$ Yang-Mills theory consists of three kinds of supermultiplets.
Two of them, the vector multiplet $(A_\mu,\lambda_1,\lambda_2,{\cal W})$ and
the adjoint hypermultiplet $({\cal Z},\wt{\cal Z},\chi,\wt\chi)$, belong to
the adjoint representation of the gauge group.
The three adjoint scalar fields $\cal Z$, $\wt{\cal Z}$ and $\cal W$
represent the fluctuations of the D3-branes along the coordinates $Z$, $\wt Z$ and $W$,
respectively.
We also have $N_f$ fundamental hypermultiplets, $(q,\wt q,\psi,\wt\psi)$,
belonging to the (anti-)fundamental representation
of the gauge group.

The set of rotations in the four NN-directions is identified with the Lorentz symmetry of
the gauge theory.
The rotation group corresponding to the four DN-directions is
$Sp(1)_R\times Sp(1)_H$.
The $Sp(1)_R$ symmetry is the R-symmetry of the ${\cal N}=2$ theory.
The symmetry $Sp(1)_H$ acts on the adjoint hypermultiplet.
By the two $Sp(1)$ symmetries,
scalar fields in the adjoint hypermultiplet
and (anti-)squark fields are transformed as
\begin{equation}
\left(\begin{array}{cc}{\cal Z}&\wt{\cal Z}\\-\wt{\cal Z}^\dagger&{\cal Z}^\dagger\end{array}\right)
\rightarrow
U_R\left(\begin{array}{cc}{\cal Z}&\wt{\cal Z}\\-\wt{\cal Z}^\dagger&{\cal Z}^\dagger\end{array}\right)U_H^{-1},
\quad
\left(\begin{array}{c}q\\-\wt q^\dagger\end{array}\right)
\rightarrow
U_R\left(\begin{array}{c}q\\-\wt q^\dagger\end{array}\right),
\label{sp1sp1}
\end{equation}
where $U_R\in Sp(1)_R$ and $U_H\in Sp(1)_H$.
The field $\cal W$ is invariant under $Sp(1)_R\times Sp(1)_H$.
The rotation group corresponding to the two DD-directions,
$U(1)_W$, represents the phase rotation of the scalar field $\cal W$.
There is another global symmetry, $U(N_f)$,
which is a gauge symmetry on the D7-branes
and acts on the fundamental hypermultiplets.
We define two $U(1)$ charges, $J_R$ and $J_H$, associated with the Cartan part
of $Sp(1)_R$ and $Sp(1)_H$, respectively.
Furthermore, we define $J$ and $\wt J$ by
\begin{equation}
J=J_R+J_H,\quad
\wt J=J_R-J_H.
\end{equation}
We also define the $U(1)_W$ charge $J_W$.
These $U(1)$ charges and spacetime spins
of each field are summarized in
Tables \ref{charges33vec.tbl}-\ref{charges37.tbl}.
\begin{table}[htb]
\caption{Charges of the fields belonging to
a vector multiplet. $\lambda_1$ and $\lambda_2$ forms an $Sp(1)_R$ doublet.}
\label{charges33vec.tbl}
\begin{center}
\begin{tabular}{ccccc}
\hline
\hline
       & spin & $J$ & $\wt J$ & $J_W$ \\
\hline
$A_\mu$ & $(2_L,2_R)$ & $0$ & $0$ & $0$ \\
$\lambda_1$ & $2_L$ & $1/2$ & $1/2$ & $1/2$ \\
$\lambda_2$ & $2_L$ & $-1/2$ & $-1/2$ & $1/2$ \\
$\cal W$ & $1$  & $0$ & $0$ & $1$ \\
\hline
\end{tabular}
\end{center}
\end{table}
\begin{table}[htb]
\caption{Charges of the fields belonging to
an adjoint hyper-multiplet.
The two complex bosons belong to ${\bf2}\times{\bf2}$ of $Sp(1)_R\times Sp(1)_H$,
and the two fermions belong to an $Sp(1)_H$ doublet.}
\label{charges33hyp.tbl}
\begin{center}
\begin{tabular}{ccccc}
\hline
\hline
       & spin & $J$ & $\wt J$ & $J_W$ \\
\hline
$\cal Z$ & $1$ & $1$ & $0$ & $0$ \\
$\chi$ & $2_L$ & $1/2$ & $-1/2$ & $-1/2$ \\
$\wt{\cal Z}$ & $1$ & $0$ & $1$ & $0$ \\
$\wt\chi$ & $2_L$ & $-1/2$ & $1/2$ & $-1/2$ \\
\hline
\end{tabular}
\end{center}
\end{table}
\begin{table}[htb]
\caption{Charges of the fields belonging to
a fundamental hyper-multiplet.
The squarks $q$ and $\wt q$ belong to $\bf 2$ of
$Sp(1)_R$.}
\label{charges37.tbl}
\begin{center}
\begin{tabular}{ccccccc}
\hline
\hline
       & spin & $J$ & $\wt J$ & $J_W$ & $ U(N_f)$ & $ U(N_c)$ \\
\hline
$q$       & $1$   & $1/2$ & $1/2$ & $0$ & $\ol N_f$ & $N_c$ \\
$\psi$    & $2_L$ & $0$   & $0$   & $-1/2$ & & \\
$\wt q$   & $1$   & $1/2$ & $1/2$ & $0$ & $N_f$ & $\ol N_c$ \\
$\wt\psi$ & $2_L$ & $0$   & $0$   & $-1/2$ & & \\
\hline
\end{tabular}
\end{center}
\end{table}

The propagators for the adjoint scalar fields are
\begin{equation}
\langle{\cal Z}^i{}_j(x){\cal Z}^{\dagger k}{}_l(y)\rangle
=\langle\wt{\cal Z}^i{}_j(x)\wt{\cal Z}^{\dagger k}{}_l(y)\rangle
=\langle{\cal W}^i{}_j(x){\cal W}^{\dagger k}{}_l(y)\rangle
=\delta^i_l\delta^k_jD(x-y),
\end{equation}
where $i$, $j$, $k$ and $l$ are the color indices.
We almost always omit the color and flavor indices in this paper.
The propagators for squarks are
\begin{equation}
\langle q_A^i(x)q^{\dagger B}_j(y)\rangle=
\langle\wt q_A^{\dagger i}(x)\wt q^B_j(y)\rangle=\delta^B_A\delta^i_jD(x-y),
\end{equation}
where $A$ and $B$ are the flavor indices.
The function $D(x)$ in these propagators is defined by
\begin{equation}
D(x)=\frac{1}{4\pi^2|x|^2}.
\end{equation}

\subsection{Gauge singlet operators}
We now consider gauge invariant operators in the ${\cal N}=2$ theory
with large R-charges $J$.
If the theory is conformal, the conformal dimension $\Delta$ of
operators is bounded below due to the BPS bound as
\begin{equation}
\Delta\geq|J|+|\wt J|.
\end{equation}
We consider only operators with small $\Delta-J$.
The values of $\Delta-J$ for fields in the theory are
given in Table \ref{d-j.tbl}.
\begin{table}[htb]
\caption{$\Delta-J$ of each field}
\label{d-j.tbl}
\begin{center}
\begin{tabular}{clll}
\hline
\hline
$\Delta-J$ & vector mult. & adj. hyper mult. & fund. hyper mult. \\
\hline
$0$ & & ${\cal Z}$ \\
\hline
$1/2$ & & & $q$, $\wt q$ \\
\hline
$1$ & $A_\mu$, $\lambda_1$, $\lambda_2^\dagger$, $\cal W$, ${\cal W}^\dagger$
& $\chi$, $\wt\chi^\dagger$, ${\cal Z}'$, ${\cal Z}{'}^\dagger$ \\
\hline
$3/2$ & & & $q^\dagger$, $\wt q^\dagger$, $\psi$, $\psi^\dagger$, $\wt\psi$, $\wt\psi^\dagger$ \\
\hline
$2$ & $\lambda_1^\dagger$, $\lambda_2$ & ${\cal Z}^\dagger$, $\chi^\dagger$, $\wt\chi$ \\
\hline
\end{tabular}
\end{center}
\end{table}
At tree level, $\Delta-J$ for a composite operator is given by
the sum of the $\Delta-J$ for the constituent fields.
We consider non-trace operators consisting of a
large number of adjoint scalar fields,
one fundamental scalar field and one anti-fundamental scalar field.
Because for fields in the fundamental and the anti-fundamental representations
the relation $\Delta-J\geq1/2$ holds,
for any non-trace operator,
we have $\Delta_{\rm tree}-J\geq1$.

In this paper, we consider non-trace operators
for which $\Delta_{\rm tree}-J=1$ or $2$ that consist of only scalar fields.
The unique non-trace operator with $\Delta_{\rm tree}-J=1$ is
\begin{equation}
O=\wt q{\cal Z}^{J-1}q.
\label{qZq}
\end{equation}
In order to determine the conformal dimension of this operator,
we need to compute its two-point function.
The tree amplitude is given by
\begin{equation}
\langle O,O\rangle
\equiv\langle O^\dagger(x)O(0)\rangle
=N_c^J D^{J+1}(x)
\end{equation}
Because the operator $O$ is chiral,
it saturates the BPS bound $\Delta-J-\wt J\geq0$
and its anomalous dimension undergoes no the quantum corrections.
More precisely, this is correct only if the theory is conformal.
This is the case only for the $Sp(N_c/2)$ theory with $N_f=8$.
However, at the order $\eta=N_cg_{\rm YM}^2$, we can easily
show that the quantum correction in the $U(N_c)$ theory is proportional
to that in the $Sp(N_c/2)$ theory and is independent of $N_f$.
Therefore, we conclude that there is no quantum correction
to the anomalous dimension of the operator $O$
at order $\eta$, even in the $U(N_c)$ theory.

In Ref.\citen{open}, it is suggested that
the operator $O$ corresponds to the ground state of 7-7 open strings
in a PP-wave background.
Open strings in the PP-wave background have only one ground state,
while strings in flat Minkowski space have sixteen degenerate
massless ground states.
If we introduce a non-vanishing $\mu$-parameter,
the degeneracy of the ground states is broken, and
only state remains as the ground state.
We can determine which of the massless ground states in the flat background
becomes the ground state in the PP-wave background
when we introduce a non-vanishing $\mu$-parameter
by comparing the charges of strings in both backgrounds.
In Ref.\citen{open}, it is shown that the open string ground state
in the PP-wave background possesses $\wt J=1$.
The unique such state in the massless spectrum of
an open string in the flat background is
$b^{\wt Z}_{-1/2}|0\rangle$.\footnote{In this paper we use the RNS formalism,
in which
$|0\rangle$ is the tachyonic ground state in the NS-sector, and
$b^\mu_{-1/2}$ is one of the oscillators of the worldsheet fermion $\psi^\mu$.}
Therefore, we should identify this state with the ground state in the PP-wave
background, and following the idea presented in Ref.\citen{open},
we can surmise the correspondence
\begin{eqnarray}
b_{-1/2}^{\wt Z}|0\rangle\quad\leftrightarrow\quad O.
\label{corresp1}
\end{eqnarray}

Operators with $\Delta_{\rm tree}-J=2$ are constructed by
replacing one of the constituent fields in $O^{(J+1)}\equiv \wt q{\cal Z}^Jq$ by another field.
We label the $J+2$ constituent fields of $O^{(J+1)}$ by $0,1,\cdots,J+1$
from right to left.
Let $O^X_k$ denote the operator obtained by exchanging the $k$-th fields in $O^{(J+1)}$ by a scalar field $X$.
Four kinds of operators are obtained by replacing one $\cal Z$ in $O^{(J+1)}$
by $\wt{\cal Z}$, $\cal W$ or their hermitian conjugates:
\begin{equation}
O^X_k=\wt q{\cal Z}^{J-k}X{\cal Z}^{k-1}q,\quad
(X=\wt{\cal Z},\wt{\cal Z}^\dagger,{\cal W},{\cal W}^\dagger)
\label{inserted}
\end{equation}
where $1\leq k\leq J$.
In addition to these operators,
there are two operators obtained by replacing one of the squarks
by an anti-squark:
\begin{equation}
O^{\wt q^\dagger}_0=\wt q{\cal Z}^J\wt q^\dagger,\quad
O^{q^\dagger}_{J+1}=q^\dagger{\cal Z}^Jq.
\label{oqbar}
\end{equation}

To establish the correspondence between operators and string states,
we need to compute two-point functions among these operators.\cite{BMN,open}
As a result, we obtain a $(4k+2)\times(4k+2)$ matrix, which is often called
the ``string-bit Hamiltonian.''
By diagonalizing the matrix, we obtain the `eigen-operators' with
well-defined conformal dimensions.
We identify each of these operators with each string state excited by one bosonic oscillator.

At order $\eta$, there are three types of non-vanishing two-point functions.
Two of them are the diagonal components $\langle O_k^X,O_k^X\rangle$
and the hopping components $\langle O_k^X,O_{k\pm1}^X\rangle$.
In addition to these,
the operators $O^{\wt q^\dagger}_0$ and $O^{q^\dagger}_{J+1}$ couple to
$O^{\wt{\cal Z}^\dagger}_1$ and $O^{\wt{\cal Z}^\dagger}_J$, respectively.
Therefore, it is convenient to define
\begin{equation}
O^{\wt{\cal Z}^\dagger}_0\equiv O^{q^\dagger}_0,\quad
O^{\wt{\cal Z}^\dagger}_{J+1}\equiv-O^{\wt q^\dagger}_{J+1}.
\end{equation}
Computing planar diagrams with one four-point interaction,
we obtain the two-point functions
\begin{equation}
\langle O^X_k,O^X_l\rangle
={\cal A}_{\rm tree}\left(\delta_{k,l}+M_{k,l}^X\frac{\eta}{4\pi^2}\log\frac{\epsilon}{|x|}\right),
\end{equation}
where $X$ represents one of $\wt{\cal Z}$, $\wt{\cal Z}^\dagger$, $\cal W$ and ${\cal W}^\dagger$,
and ${\cal A}_{\rm tree}$ is defined by
\begin{equation}
{\cal A}_{\rm tree}=\langle O^{(J+1)},O^{(J+1)}\rangle
=N_c^{J+1}D^{J+2}(x).
\end{equation}
The matrices $M_{k,l}^X$ are given by
\begin{equation}
M^{\wt{\cal Z}}_{k,l}=\left(\begin{array}{ccccc}
-2 &  2 \\
 2 & -4 & 2 \\
   &  2 & \ddots & \ddots \\
   &    & \ddots & -4 & 2 \\
   &    &        &  2 & -2
\end{array}\right),
\quad(k,l=1,2,\cdots,J)
\end{equation}
\begin{equation}
M^{\wt{\cal Z}^\dagger}_{k,l}=\left(\begin{array}{ccccc}
-2 &  2 && \\
 2 & -4 & 2 & \\
   &  2 & \ddots & \ddots \\
   &    & \ddots & -4 & 2 \\
   &    &        &  2 & -2
\end{array}\right),
\quad(k,l=0,1,\cdots,J+1)
\end{equation}
\begin{equation}
M^{\cal W}_{k,l}=M^{{\cal W}^\dagger}_{k,l}
=\left(\begin{array}{ccccc}
-4 &  2 \\
 2 & -4 & 2 \\
   &  2 & \ddots & \ddots \\
   &    & \ddots & -4 & 2 \\
   &    &        &  2 & -4
\end{array}\right).
\quad(k,l=1,2,\cdots,J)
\end{equation}

At leading order in $1/J$, the small eigenvalues of these matrices are
given by
\begin{equation}
\rho=-\frac{2n^2\pi^2}{J^2},\quad
\left\{\begin{array}{l}
\mbox{$n=0,1,2,\cdots$ for $X=\wt{\cal Z},\wt{\cal Z}^\dagger$},\\
\mbox{$n=1,2,3,\cdots$ for $X={\cal W},{\cal W}^\dagger$}.
\end{array}\right.
\end{equation}
Correspondingly, we have
\begin{equation}
\Delta-J
=2-\frac{\eta\rho}{8\pi^2}
=2+\frac{\eta}{4J^2}n^2.
\label{dmj}
\end{equation}
Following Ref.\citen{BMN}, this should be identified with the
light-cone Hamiltonian of open strings in the PP-wave background.
The spectrum is given by\cite{open}
\begin{equation}
P^-=1+\sum_{l=0}^\infty N_l\sqrt{1+\frac{\eta}{2J^2}l^2},
\label{ppspec}
\end{equation}
where we have normalized the energy so that the ground state energy is $1$
and have used the relation $R^4=4\pi N_cg_{\rm str}\alpha{'}^2$
and $g_{\rm str}=2\pi g_{\rm YM}^2$
to rewrite the spectrum in terms of the gauge theory parameters.
We see that (\ref{dmj}) coincides with (\ref{ppspec}) up to order
$\eta^2$ when we set $N_l=\delta_{n,l}$.
We can easily conjecture the correspondence between
string excited states and `eigen-operators' of the matrices
as follows:
\begin{eqnarray}
a_{-n}^{\wt Z}b_{-1/2}^{\wt Z}|0\rangle&\quad\leftrightarrow\quad&
\sum_{k=1}^JO^{\wt{\cal Z}}_k\cos\frac{\pi n(k-1/2)}{J},
\label{corresp2}\\
a_{-n}^{\wt Z^\dagger}b_{-1/2}^{\wt Z}|0\rangle&\quad\leftrightarrow\quad&
\sum_{k=0}^{J+1}O^{\wt{\cal Z}^\dagger}_k\cos\frac{\pi n(k+1/2)}{J+2},
\label{corresp3}\\
a_{-n}^Wb_{-1/2}^{\wt Z}|0\rangle&\quad\leftrightarrow\quad&
\sum_{k=1}^JO^{\cal W}_k\sin\frac{\pi nk}{J+1},
\label{corresp4}\\
a_{-n}^{W^\dagger}b_{-1/2}^{\wt Z}|0\rangle&\quad\leftrightarrow\quad&
\sum_{k=1}^JO^{{\cal W}^\dagger}_k\sin\frac{\pi nk}{J+1}.
\label{corresp5}
\end{eqnarray}

The boundary conditions for the wave functions of the operators in Eqs.(\ref{corresp2})--(\ref{corresp5}) can easily be guessed
using the symmetry argument.
Because $Sp(1)_H$ and $Sp(1)_R$ are the symmetries of the Lagrangian,
the operators obtained from $O^{(J+1)}$ by the rotation (\ref{sp1sp1}) also
undergoes no quantum corrections.
In fact, such $Sp(1)$ rotations generate the operators
in (\ref{corresp2}) and (\ref{corresp3}) with $n=0$:
\begin{equation}
\delta_{Sp(1)_H}O^{(J+1)}
=\sum_{k=1}^JO^{\wt{\cal Z}}_k,\quad
\delta_{Sp(1)_R}O^{(J+1)}
=\sum_{k=1}^JO^{\wt{\cal Z}^\dagger}_k
  +O^{\wt q^\dagger}_0-O^{q^\dagger}_{J+1}.
\end{equation}
This strongly suggests that the `boundary conditions' for the wave functions
of these operators are Neumann.
However, we cannot obtain operators with a $\cal W$ or ${\cal W}^\dagger$ insertion
by these rotations.
This is consistent with the fact that the boundary conditions for operators with $\cal W$ and ${\cal W}$
insertions are Dirichlet.

\subsection{Orientifold projection}\label{oripro.ssec}
The $Sp(N_c/2)$ theory realized on D3-branes in the D7-O7 background
is constructed with a ${\bf Z}_2$ orientifold projection
from the $U(N_c)$ theory.
We divide all the fields into ${\bf Z}_2$ even and odd parts as
\begin{eqnarray}
{\cal J}{\cal Z}_{(\pm)}&=&\frac{1}{\sqrt2}({\cal J}{\cal Z}\mp({\cal J}{\cal Z})^T),\label{Zpm}\\
{\cal J}\wt{\cal Z}_{(\pm)}&=&\frac{1}{\sqrt2}({\cal J}\wt{\cal Z}\mp({\cal J}\wt{\cal Z})^T),\\
{\cal J}{\cal W}_{(\pm)}&=&\frac{1}{\sqrt2}({\cal J}{\cal W}\pm({\cal J}{\cal W})^T),\\
{\cal J}q_{(\pm)}&=&\frac{1}{\sqrt2}({\cal J}q\mp\wt q^T),
\end{eqnarray}
where ${\cal J}$ is the anti-symmetric matrix used to define the $Sp(N_c/2)$
group.
If we remove all the ${\bf Z}_2$ odd fields from the
action, we obtain the action of the $Sp(N_c/2)$ theory.

The gauge invariant operators in the $Sp(N_c/2)$ theory are defined by replacing all the constituent fields
in the operators in the $U(N_c)$ theory by the ${\bf Z}_2$ invariant fields.
The unique operator with $\Delta_{\rm tree}-J=1$ is
\begin{equation}
O=q_{(+)}^T{\cal J}{\cal Z}_{(+)}^{J-1}q_{(+)}.
\end{equation}
The operators for which $\Delta_{\rm tree}-J=2$ are defined by
\begin{equation}
O^X_k=q_{(+)}^T{\cal J}{\cal Z}_{(+)}^{J-k}X_{(+)}{\cal Z}_{(+)}^{k-1}q_{(+)},\quad
X=
\wt {\cal Z},
\wt {\cal Z}^\dagger,
{\cal W},
{\cal W}^\dagger.
\end{equation}
\begin{equation}
O^{\wt{\cal Z}^\dagger}_0=q^T{\cal J}{\cal Z}^J{\cal J}^{-1}q^\dagger,\quad
O^{\wt{\cal Z}^\dagger}_{J+1}=q^\dagger{\cal Z}^Jq.
\end{equation}
These operators have two flavor indices.
Because of the ${\bf Z}_2$ symmetry, only half of these operators are independent.
They satisfy
\begin{equation}
O_{AB}=-O_{BA},
\end{equation}
and
\begin{equation}
(O^X_k)_{AB}=\pm (O^X_{J-k})_{BA},\quad
\mbox{$+$ for $X=\wt{\cal Z}, \wt{\cal Z}^\dagger$, $-$ for $X={\cal W},{\cal W}^\dagger$.}
\end{equation}
These relations impliy that the two flavor indices of the BMN operators
in (\ref{corresp2})--(\ref{corresp5}) are symmetric when $n$ is even and anti-symmetric when $n$ is odd.
This is consistent with the orientifold projections of the string states
on the left-hand sides of Eqs.(\ref{corresp2})--(\ref{corresp5}).\cite{open}

\section{Coupling to 7-7 strings}\label{string.sec}
\subsection{Ground state}\label{ground.ssec}
In this section, we reproduce the operator-open string state
correspondence given in (\ref{corresp1}) and (\ref{corresp2})--(\ref{corresp5})
by the relation (\ref{circ-on}).
We define the coupling $\langle B|n\rangle$ by
a disk amplitude.
The boundary of the disk is divided into two parts, the D7- and D3-boundaries.
The vertex operator of the state $|n\rangle$ is inserted on
the D7-boundary.
The D3-brane boundary interacts with the scalar fields
on D3-branes via the interaction term
\begin{equation}
S_{\rm int}=g_{\rm str}\int_{\mbox{D3-boundary}}\sum_XV_Xd\tau+g_{\rm str}\sum_QV_Q.
\label{Sint}
\end{equation}
For simplicity, we assume that
the scalar fields on D3-branes are constant
and that the fermions and the vector field vanish.
The first term in (\ref{Sint}) represents the interaction with the adjoint scalar fields on the D3-branes.
There are six adjoint scalars, and their corresponding vertex operators
in picture $0$ are\footnote{In this section we use the convention in which
the scalar field $X^\mu$ is anti-hermitian to avoid the appearance
of many imaginary units$i$.}
\begin{eqnarray}
&&
V_{\cal Z}={\cal Z}\partial Z^\dagger,\quad
V_{\wt{\cal Z}}=\wt{\cal Z}\partial\wt Z^\dagger,\quad
V_{\cal W}={\cal W}\partial W^\dagger,\nonumber\\
&&
V_{{\cal Z}^\dagger}={\cal Z}^\dagger\partial Z,\quad
V_{\wt{\cal Z}^\dagger}=\wt{\cal Z}^\dagger\partial\wt Z,\quad
V_{{\cal W}^\dagger}={\cal W}^\dagger\partial W.
\end{eqnarray}
We include the wave functions of fields in the definition of the vertex operators.
The operator $V_Q$ in (\ref{Sint}) is the vertex operator for squarks and anti-squarks.
Here, $Q$ represents one of $q$, $\wt q$, $q^\dagger$ and $\wt q^\dagger$.
They are inserted at points dividing D3- and D7-boundaries.
For each squark and anti-squark,
the vertex operator in picture $-1$ is given by
\begin{equation}
V_{\wt q}=\wt qe^{i\phi}\sigma S^{--},\quad
V_{\wt q^\dagger}=\wt q^\dagger e^{i\phi}\sigma S^{++},\quad
V_q=qe^{i\phi}\sigma S^{--},\quad
V_{q^\dagger}=q^\dagger e^{i\phi}\sigma S^{++}.
\label{Vq}
\end{equation}
We call these `twisting vertices'.
Here, $S^{\pm\pm}$ represents the spin operator in the DN-directions.
The left and right superscripts represent the signs of $J$ and $\wt J$, respectively.
For example, $S^{--}$ has $J=\wt J=-1/2$.
The operator $\phi$ is one of the bosonized superconformal ghost fields,
and the factor $e^{i\phi}$ implies that the vertices are in picture $-1$.
The operator $\sigma$ is the twist operator for four
DN-directions.
Explicitly, $\sigma$ is defined as an operator satisfying\cite{orbi1,orbi2}
\begin{equation}
\partial X^i(z_1)\sigma(z_2)=\frac{1}{(z_1-z_2)^{1/2}}\tau^i(z_2)+{\cal O}((z_1-z_2)^{1/2}).\quad
(i=4,5,6,7)
\label{twistope}
\end{equation}
From the OPE (\ref{twistope}), we can uniquely determine the two-point functions
of the boson fields $X^i$ in the background with two twist operators
inserted at $z=0$ and $z=\infty$ as
\begin{equation}
\langle\partial X^i(z_1)\partial X^j(z_2)\rangle=
\delta^{ij}\frac{1}{2}\left(\frac{\sqrt{z_1}}{\sqrt{z_2}}
                +\frac{\sqrt{z_2}}{\sqrt{z_1}}\right)
\frac{1}{(z_1-z_2)^2}.\quad
(i,j=4,5,6,7)
\label{twistprop}
\end{equation}
From this two-point function,
we can show that the conformal dimension of the twist operator
$\sigma$ is $1/4$.

First, let us consider the ground state $|\zeta\rangle=\zeta_\mu b^\mu_{-1/2}|0\rangle$ with angular momentum $J$ in the $Z$-plane.
We need to construct a vertex operator for this state.
This can be done as follows.
We start from the vertex operator for the plane wave of the massless vector state,
\begin{equation}
V_{\rm ground}=\zeta_\mu(\partial X^\mu-\psi^\mu k_\nu\psi^\nu)e^{k_\lambda X^\lambda}.
\end{equation}
For this state to be physical,
the momentum $k$ and the polarization $\zeta$ must satisfy
$k^2=k\cdot\zeta=0$.
We define the light-cone coordinate and the Minkowski metric as
\begin{equation}
A^\pm=\frac{1}{\sqrt2}(A^8\pm A^9),\quad
A\cdot B=A^+B^-+A^-B^++A^iB^i,
\end{equation}
and we take the components of the vectors as
\begin{equation}
(k^i,k^+,k^-)=(0,k^+,0),\quad
(\zeta^i,\zeta^+,\zeta^-)=(\zeta^i,0,0).\label{kzcond}
\end{equation}
We consider the time direction to be $X^9$, temporarily.
For the vectors (\ref{kzcond}),
the vertex is
\begin{equation}
V_{\rm ground}=\zeta_i(\partial X^i-\psi^i k^-\psi^+)e^{k^-X^+}.
\label{V56}
\end{equation}
We can check that this vertex represents a physical state by
computing the OPE of this vertex with the BRS current.
For plane waves, the momenta $k^\pm$ and the fields $iX^\pm$ are real.
However, we do not have to use this fact to show that (\ref{V56}) is a physical vertex.
To prove that the vertex operator is physical,
we need to demonstrate OPEs like
\begin{equation}
\partial X^+(z_1)\partial X^-(z_2)=\frac{1}{(z_1-z_2)^2}.
\label{OPExpm}
\end{equation}
Therefore, we can relax the condition that the momenta $k^\pm$ and the field
$iX^\pm$ be real without spoiling the physical nature of the vertex.
We replace the light-cone components of
the vectors by complex variables as follows:
\begin{equation}
X^+\rightarrow Z,\quad
X^-\rightarrow Z^\dagger,\quad
k^+\rightarrow k^Z,\quad
k^-\rightarrow k^{Z^\dagger}.
\label{Wickreplace}
\end{equation}
Now, instead of the condition that variables be real, the new variables satisfy the relations
$(k^Z)^\dagger=k^{Z^\dagger}$ and $(iZ)^\dagger=iZ^\dagger$.
This replacement is equivalent to the Wick rotation $X^9\rightarrow iX^9$,
and the `time' direction $X^9$ becomes space-like.
Because the OPE (\ref{OPExpm}) is left invariant under this replacement,
it maps physical vertices to physical vertices.
Therefore, we obtain
\begin{equation}
V_{\rm ground}=\zeta_i(\partial X^i-\psi^i k^{Z^\dagger}\psi^Z)e^{k^{Z^\dagger}Z}.
\label{wickedV}
\end{equation}
This is a superposition of states with various positive angular momenta.
Because the BRS charge is rotationally invariant,
each part carrying a distinct angular momentum is separately a physical vertex.
If we pick out the terms with angular momentum $J$ from
(\ref{wickedV}), we obtain the circular wave state vertex operator
\begin{equation}
V^J_{\rm ground}
=\frac{1}{J!}\zeta_i(\partial X^iZ^J-J\psi^i\psi^ZZ^{J-1}).
\label{masslesscir}
\end{equation}
[We have removed the overall numerical factor $(k^{Z^\dagger})^J$.]
Although the norm of this state is not well-defined,
this is a physical vertex by construction.

We now consider the disk amplitude with $V_{\rm ground}^J$ inserted on the
D7-boundary.
Because $V_{\rm ground}^J$ includes
at least $J-1$ fields $Z$,
we need to insert at least $J-1$ vertices $V_{\cal Z}$ on the D3-boundary.
As the number of the vertices increases,
the power of $g_{\rm str}$ also increases.
Because we are assuming small string coupling constant $g_{\rm str}\ll1$,
the dominant contribution comes from the amplitude with the smallest number
of the insertions.
Therefore, we consider the case that
$J-1$ vertices $V_{\cal Z}$ are inserted on the boundary.
In this case, only the second term of $V^J_{\rm ground}$ contributes to
the amplitude.

\begin{figure}[htb]
\centerline{\includegraphics{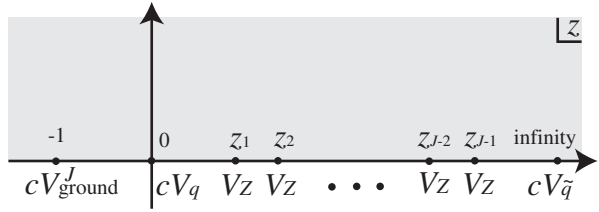}}
\caption{Worldsheet representing the amplitude $\langle B|\zeta\rangle$.}
\label{uhp.eps}
\end{figure}

When determining which term contributes to the amplitude,
charge conservation laws are useful.
We can separate the amplitude into several sectors.
Correspondingly, the charges $J$ and $\wt J$
are divided into $J_B$ and $\wt J_B$ carried by the boson fields
$X^\mu$ and $J_F$ and $\wt J_F$ carried by the fermions $\psi^\mu$.
These four charges are conserved separately.
Because the vertex operators for the adjoint scalar fields do not contain
fermions,
the angular momentum $J_F=1$ carried by the second term
of the vertex operator $V_{\rm ground}^J$ should be cancelled by
the vertex operators of the (anti-)squarks (\ref{Vq}).
Because the squark vertex and the anti-squark vertex have $J_F=-1/2$ and $J_F=1/2$,
respectively, both twisting vertices should be squark vertices.
Altogether, on the boundary of the disk,
two squark vertices, $V_q$ and $V_{\wt q}$, $J-1$ vertices $V_{\cal Z}$
and $V_{\rm ground}^J$ are inserted.
We use the upper half plane to represent an open string worldsheet,
and we place one squark vertex at each $z=0$ and $z=\infty$.
The positive and the negative parts of the real axis represent
the D3- and D7-boundaries, respectively.
The $J-1$ vertices $V_{\cal Z}$ are inserted on the positive real axis
at positions $z_i$ satisfying
\begin{equation}
0<z_1<z_2<\cdots<z_{J-2}<z_{J-1}<\infty.
\label{zzz}
\end{equation}
We place the 7-7 string vertex at $z=-1$
(see Fig. \ref{uhp.eps}).

\begin{figure}
\centerline{\includegraphics{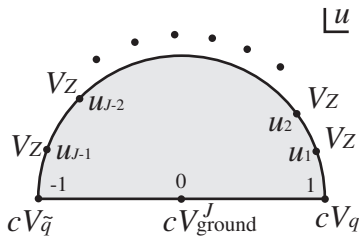}}
\caption{The upper half plane in the $z$-coordinate (Fig. \ref{uhp.eps})
is mapped into the upper half unit disk in the $u$-coordinate.}
\label{hd1.eps}
\end{figure}

We fix the residual ${\rm SL}(2,{\bf R})$ diffeomorphism
by fixing the coordinates of the vertices of $q$, $\wt q$ and the 7-7 string, and
we have the following expression for the disk amplitude:
\begin{eqnarray}
\langle B|\zeta\rangle
&=&g_{\rm str}^J\int_0^\infty d^{J-1}z
\Big\langle
\ cV_{\wt q}(\infty)
\ V_Z(z_{J-1})
\ \cdots
\ V_Z(z_1)
\ cV_q(0)
\ cV^J_{\rm ground}(-1)
\ \Big\rangle\nonumber\\
&&+{\cal O}(g_{\rm str}^{J+1}).
\label{BZ62}
\end{eqnarray}
The integral $\int_0^\infty d^{J-1}z$ in (\ref{BZ62}) is
an integral over $J-1$ variables $z_k$ satisfying (\ref{zzz}):
\begin{equation}
\int_0^\infty d^{J-1}z=\int_0^\infty dz_{J-1}
\int_0^{z_{J-1}}dz_{J-2}
\cdots
\int_0^{z_3}dz_2
\int_0^{z_2}dz_1.
\label{dzdef}
\end{equation}
We normalize the correlation functions in each sector according to
\begin{equation}
\langle c(\infty)c(0)c(-1)\rangle
=\langle e^{i\phi}(\infty)e^{i\phi}(0)\rangle
=\langle\sigma(\infty)\sigma(0)\rangle
=\langle S^{++}(\infty)S^{--}(0)\rangle
=1.
\label{znorm}
\end{equation}
As we mentioned above,
due to conservation of $J_F$,
only the second term
in the vertex (\ref{masslesscir}) contributes to the amplitude.
If we first evaluate the $\psi^\mu$ sector, we have
\begin{equation}
\langle S^{--}(\infty)S^{--}(0)V_{\rm ground}^J(-1)\rangle
=\zeta_{\wt Z}\frac{[Z(-1)]^{J-1}}{(J-1)!}.
\end{equation}
After using the normalization (\ref{znorm}),
we are left with the product of $J-1$ bosonic two-point functions and
the integral over $z_k$ as
\begin{equation}
\langle B|\zeta\rangle=g_{\rm str}^J\zeta_{\wt Z}(\wt q{\cal Z}^{J-1}q)
\int d^{J-1}z
\prod_{k=1}^{J-1}
\langle\partial Z^\dagger(z_k)Z(-1)\rangle_{\rm twisted}
+{\cal O}(g_{\rm str}^{J+1}).
\label{amp2}
\end{equation}
where $\langle\cdots\rangle$ represents the correlation function
in the background with the twist operators $\sigma(0)$ and $\sigma(\infty)$
given in (\ref{twistprop}).
This correlation function is a double-valued function, due to the twist operators.
To treat such a propagator, it is convenient to use the covering coordinate
$w=z^{1/2}$ on the Riemann surface.
Under the coordinate transformation, the propagator
takes a simple form, without square root,
\begin{equation}
\langle\partial Z^\dagger(w_1)\partial Z(w_2)\rangle
=\frac{1}{(w_1-w_2)^2}+\frac{1}{(w_1+w_2)^2}.
\end{equation}
Integrating this with respect to $w_2$, we have
\begin{equation}
\langle\partial Z^\dagger(w_1)Z(w_2)\rangle
=\frac{1}{(w_1-w_2)}-\frac{1}{(w_1+w_2)}
=\frac{2w_2}{w_1^2-w_2^2},
\end{equation}
and each integral in (\ref{amp2}) can be rewritten as
\begin{equation}
\int\langle\partial Z^\dagger(z_k)Z(-1)\rangle dz_k
=\int\langle\partial Z^\dagger(w_k)Z(i)\rangle dw_k
=\int\frac{2idw_k}{w_k^2+1}.
\label{zandw}
\end{equation}
This is further simplified if we introduce a coordinate $u$ defined by
\begin{equation}
w=i\frac{1-u}{1+u}.
\end{equation}
Using this coordinate, (\ref{zandw}) is rewritten as
\begin{equation}
\int\langle\partial Z^\dagger(z_k)Z(-1)\rangle dz_k
=\int\frac{du_k}{u_k}.
\end{equation}
By this coordinate transformation,
the worldsheet (the upper half plane in the $z$ coordinate)
is mapped into the upper half unit disk $\{u|\Im u\geq0,|u|\leq1\}$
in the $u$-plane, and
the vertices of the adjoint scalar fields are inserted on the arc
of the half disk
(see Fig. \ref{hd1.eps}).
We have the following expression for the amplitude in this coordinate:
\begin{equation}
\langle B|\zeta\rangle=g_{\rm str}^J\zeta_{\wt Z}(\wt q{\cal Z}^{J-1}q)
\int\left[\frac{du}{u}\right]^{J-1}1
+{\cal O}(g_{\rm str}^{J+1}).
\end{equation}
Here, $\int[du/u]^{J-1}$ represents integration
with respect to the $J-1$ variables $u_k$ on the arc of the
half disk.
We can easily carry out this integral
by introducing angular variables $\theta_k$ defined by $u_k=e^{i\theta_k}$.
We have
\begin{equation}
I_{J-1}\equiv\int\left[\frac{du}{u}\right]^{J-1}1
=i^{J-1}\int_0^\pi d^{J-1}\theta
=\frac{(\pi i)^{J-1}}{(J-1)!},
\end{equation}
where $\int_0^\pi d^{J-1}\theta$ is defined similarly to (\ref{dzdef}).
We obtain the final result as
\begin{equation}
\langle B|\zeta\rangle=g_{\rm str}^JI_{J-1}\zeta_{\wt Z}O
+{\cal O}(g_{\rm str}^{J+1}).
\end{equation}
This equation implies that the state $b^{\wt Z}_{-1/2}|0\rangle$
is mapped to the gauge singlet operator $O$ and
the other ground states are mapped to operators
that do not consist only of scalar fields.
This is consistent with the correspondence (\ref{corresp1}).

\subsection{Excited states}\label{excited.ssec}
Next, we consider the excited open string state
$|n,\xi,\zeta\rangle=\xi_\mu a_{-n}^\mu\zeta_\nu b_{-1/2}^\nu|0\rangle$
with angular momentum $J$.
The vertex operator for such circular wave excited states
can be obtained with a the Wick rotation of a plane wave.
Let us start from the plane wave vertex
operator obtained from the DDF construction\cite{DDF,BF},
\begin{equation}
V_{\rm excited}(z)=\oint\frac{dw}{2\pi i} V_1(w)V_2(z),
\label{DDF}
\end{equation}
where the two operators $V_a$ ($a=1,2$) are defined by
\begin{equation}
V_1=\xi_\mu(\partial X^\mu-\psi^\mu p_\nu\psi^\nu)e^{p_\lambda X^\lambda},\quad
V_2=\zeta_\mu(\partial X^\mu-\psi^\mu k_\nu\psi^\nu)e^{k_\lambda X^\lambda}.
\end{equation}
We choose the vectors $p$, $k$, $\xi$ and $\zeta$ with the following components:
\begin{eqnarray}
&&(p^i,p^+,p^-)=(0,p^+,0),\quad
(k^i,k^+,k^-)=(0,0,k^-),
\nonumber\\&&
(\xi^i,\xi^+,\xi^-)=(\xi^i,0,0),\quad
(\zeta^i,\zeta^+,\zeta^-)=(\zeta^i,0,0).
\end{eqnarray}
The level $n$ of the excitation is determined by
\begin{equation}
p^+k^-=-n.
\end{equation}

We employ the normal order for the $X^\pm$ fields
in (\ref{DDF}) as
\begin{equation}
e^{p^+X^-(w)}e^{k^-X^+(z)}=\frac{1}{(w-z)^n}:e^{p^+X^-(w)}e^{k^-X^+(z)}:.
\label{normalorder}
\end{equation}
After this regularization, we carry out the Wick rotation through the replacement
(\ref{Wickreplace}).
Then, we have the following expression for the excited vertex operator:
\begin{eqnarray}
V_{\rm excited}(z)&=&\oint\frac{dw}{2\pi i}\frac{1}{(w-z)^n}\ :e^{p^ZZ^\dagger(w)}e^{k^{Z^\dagger}Z(z)}:
\nonumber\\&&\quad\times\quad
\xi_i(\partial X^i(w)-p^Z\psi^i(w)\psi^{Z^\dagger}(w))
\nonumber\\&&\quad\times\quad
\zeta_j(\partial X^j(z)-k^{Z^\dagger}\psi^j(z)\psi^Z(z)).
\end{eqnarray}
Note that we should carry out the Wick rotation after regularization
by the normal ordering (\ref{normalorder}), because
the operator $e^{p^ZZ^\dagger(w)}e^{k^{Z^\dagger}Z(z)}$ without
normal ordering is not well-defined.
If we pick up a part with angular momentum $J$, we obtain the following
vertex operator for a circular wave excited state:
\begin{equation}
V^J_{\rm excited}(z)=\oint\frac{dw}{2\pi i}\frac{1}{(w-z)^n}
\sum_{s=0,\pm1}:f_{J-s,n}(Z^\dagger(w),Z(z)):{\cal F}_s(w,z).
\label{VJex}
\end{equation}
Here we have removed the overall factor $(k^{Z^\dagger})^J$.
The function $f_{L,n}$ represents the wave function of the center of mass, and
${\cal F}_s$ represents the excitation of the string.
The total angular momentum $J$ is the sum of the orbital angular momentum
$J-s$ carried by $f_{J-s,n}$ and the internal spin $s$ carried by ${\cal F}_s$.
The wave function $f_{L,n}$ and the operator ${\cal F}_s$ are defined by
\begin{equation}
f_{L,n}(Z^\dagger,Z)=\sum_{k=0}^\infty\frac{(-n)^k{Z^\dagger}^kZ^{k+L}}{k!(k+L)!},
\label{wavefuncf}
\end{equation}
and
\begin{eqnarray}
{\cal F}_1(w,z)
&=&-\xi_i\partial X^i(w)
   \zeta_j\psi^j(z)\psi^Z(z),\\
{\cal F}_0(w,z)
&=&\xi_i\partial X^i(w)   \zeta_j\partial X^j(z)
   +n  \xi_i\psi^i(w)   \zeta_j\psi^j(z)   \psi^{Z^\dagger}(w)\psi^Z(z),\label{F0def}\\
{\cal F}_{-1}(w,z)
&=&n\xi_i \psi^i(w)\psi^{Z^\dagger}(w)\zeta_j \partial X^j(z),
\end{eqnarray}
respectively.

First, let us consider the spin-one term contribution.
We denote it by $\langle B|n,\xi,\zeta\rangle_{s=1}$.
Because the spin-one term
includes $J-1$ fields $Z$ in $f_{J-1,n}$ and one field $\xi_i\partial X^i$ in ${\cal F}_1$,
there should be $J-1$ vertices $V_Z$ and one vertex $V_X$ inserted on the D3-boundary,
and the non-vanishing contribution is represented by
\begin{eqnarray}
\langle B|n,\xi,\zeta\rangle_{s=1}
&=&g_{\rm str}^{J+1}\sum_{k=1}^J\int d^Ju
\Big\langle
\ cV_{\wt q}(-1)
\ V_Z(u_J)
\ \cdots
\ V_X(u_k)
\ \cdots
\nonumber\\&&
\cdots
\ V_Z(u_1)
\ cV_q(1)
\ cV_{\rm excited}^{J(s=1)}(0)
\ \Big\rangle
+{\cal O}(g^{J+2}).
\end{eqnarray}
Here we use the $u$-coordinate from the beginning.
The normalization (\ref{znorm}) in the $z$-plane is
equivalent to the following normalization in the $u$-plane
\begin{equation}
4\langle c(1)c(0)c(-1)\rangle
=\langle e^{i\phi}(1)e^{i\phi}(-1)\rangle
=\langle\sigma(1)\sigma(-1)\rangle
=\langle S^{++}(1)S^{--}(-1)\rangle
=1.
\label{unorm}
\end{equation}
The factor $4$ in the ghost sector comes from the conformal factor
$\partial z/\partial u|_{u=0}=4$.
This rescales the ghost operator at $u=0$ by $4$.
We do not have to take account of the rescaling of other operators at $u=\pm1$, because
their conformal factors cancel each other.

As we have seen in \S\ref{ground.ssec}, each pair of operators $\partial Z^\dagger$ and $Z$ contracted is replaced by $1/u$.
As a result, we obtain
\begin{eqnarray}
\langle B|n,\xi,\zeta\rangle_{s=1}
&=&\frac{1}{4}g_{\rm str}^{J+1}\sum_i\sum_{k=1}^JO^{X^i}_k\int\left[\frac{du}{u}\right]^Ju_k\oint\frac{du'}{2\pi i}\frac{1}{u{'}^n}
\nonumber\\&&
\times\langle S^{--}(1)\partial X_i(u_k){\cal F}_1(u',0) S^{--}(-1)\rangle
+{\cal O}(g_{\rm str}^{J+2})
\nonumber\\
&=&g_{\rm str}^{J+1}\sum_i\sum_{k=1}^JO^{X^i}_k\xi_j\zeta_{\wt Z}
\int\left[\frac{du}{u}\right]^Ju_k\oint\frac{du'}{2\pi i}\frac{1}{u{'}^n}
\langle \partial X^i(u_k)\partial X^j(u')\rangle
\nonumber\\
&&+{\cal O}(g_{\rm str}^{J+2}).
\label{ampex2}
\end{eqnarray}
To obtain the last expression in (\ref{ampex2}),
we have used the relation
\begin{equation}
\langle S^{--}(1)\psi^{\wt Z}(0)\psi^Z(0)S^{--}(-1)\rangle=-4.
\end{equation}
The two-point function of the scalar fields in the $u$-plane
is given by
\begin{eqnarray}
\langle\partial X(u_1)\partial X^\dagger(u_2)\rangle
&=&\frac{1}{(u_1-u_2)^2}\pm\frac{1}{(1-u_1u_2)^2}
=\sum_{n=1}^\infty n\left(\frac{1}{u_1^{n+1}}\pm u_1^{n-1}\right)u_2^{n-1},\nonumber\\&&
\left\{\begin{array}{ll}
\mbox{$+$ for $X=Z,Z^\dagger$, $\wt Z$, $\wt Z^\dagger$}, \\
\mbox{$-$ for $X=W$, $W^\dagger$.}
\end{array}\right.
\label{bbonu}
\end{eqnarray}
(Here we have assumed $|u_1|>|u_2|$ to obtain the last expansion.)
Substituting this into (\ref{ampex2}), we obtain
\begin{eqnarray}
\langle B|n,\xi,\zeta\rangle_{s=1}
&=&g_{\rm str}^{J+1}\sum_{k=1}^J\int\left[\frac{du}{u}\right]^J
\Big[2n(\xi_{\wt Z}\zeta_{\wt Z}O^{\wt{\cal Z}}_k
      +\xi_{\wt Z^\dagger}\zeta_{\wt Z}O^{\wt{\cal Z}^\dagger}_k)\cos(n\theta_k)
\nonumber\\&&
     -2in(\xi_W\zeta_{\wt Z}O^{\cal W}_k
      +\xi_{W^\dagger}\zeta_{\wt Z}O^{{\cal W}^\dagger}_k)\sin(n\theta_k)\Big]
+{\cal O}(g_{\rm str}^{J+2}),
\label{uintegral}
\end{eqnarray}
where $e^{i\theta_k}=u_k$.
We have to compute an integral of the form
\begin{equation}
I=\int^\pi_0d^J\theta f(\theta_k),
\end{equation}
where the function $f(\theta_k)$ depends on only one variable, $\theta_k$.
By integrating over the $J-1$ variables $\theta_l$ ($l\neq k$),
we have
\begin{equation}
I=\int^\pi_0d\theta_k\int^{\theta_k}_0d^{k-1}\theta\int^\pi_{\theta_k}d^{J-k}\theta f(\theta_k)
=\int^\pi_0d\theta_k\frac{\theta_k^{k-1}}{(k-1)!}\frac{(\pi-\theta_k)^{J-k}}{(J-k)!}f(\theta_k).
\label{inttk}
\end{equation}
Here, we assume that $J$ is very large and that both $k$ and $J-k$ are
of the same order of magnitude as $J$.
Then, this integral can be evaluated by the steepest descent method, and we have
\begin{equation}
I=\int^\pi_0d^J\theta f(\ol\theta_k)+{\cal O}(1/J),
\end{equation}
where $\ol\theta_k$ represents $J$ even interval points in $[0,\pi]$
defined by
\begin{equation}
\ol\theta_k=\frac{k}{J+1}\pi.
\end{equation}
Therefore, we can replace the integrated variables $\theta_k$ in (\ref{uintegral})
by $\ol\theta_k$.
With this replacement,
we obtain
\begin{eqnarray}
\langle B|n,\xi,\zeta\rangle_{s=1}
&=&2g_{\rm str}^{J+1}nI_J\sum_{k=1}^J
\Big[(\xi_{\wt Z}\zeta_{\wt Z}O^{\wt{\cal Z}}_k
      +\xi_{\wt Z^\dagger}\zeta_{\wt Z}O^{\wt{\cal Z}^\dagger}_k)\cos(n\ol\theta_k)
\nonumber\\&&\hspace{3em}
     -i(\xi_W\zeta_{\wt Z}O^{\cal W}_k
      +\xi_{W^\dagger}\zeta_{\wt Z}O^{{\cal W}^\dagger}_k)\sin(n\ol\theta_k)\Big]
\nonumber\\&&
+{\cal O}(J^{-1})+{\cal O}(g_{\rm str}^{J+2}).
\end{eqnarray}

Next, let us consider the contribution of the spin-zero term $:f_{J,n}:{\cal F}_0$.
Because the spin-one contribution is of order $g_{\rm str}^{J+1}$,
we ignore any contribution of higher order than it.
Because $:f_{J,n}:$ includes at least $J$ fields $Z$,
each of the $J$ insertions on the D3-brane boundary should be a $V_{\cal Z}$
to obtain a non-vanishing amplitude.
Because ${\cal F}_0$ has $J_F=0$, one of the two twisting operators should
be a squark, and the other should be an anti-squark.
There are two possibilities.
One is that the two twisting vertices are $V_{\wt q}(1)$ and
$V_{\wt q^\dagger}(-1)$, and
the other is that they are $V_{q^\dagger}(1)$ and $V_q(-1)$.
Let $\langle B|n,\xi,\zeta\rangle_{\wt q,\wt q^\dagger}$ and
$\langle B|n,\xi,\zeta\rangle_{q^\dagger,q}$ represent
the contributions of the former and the latter possibilities,
respectively.
Because these two can be treated analogously,
we treat only $\langle B|n,\xi,\zeta\rangle_{\wt q,\wt q^\dagger}$ in detail
and we give only the result for $\langle B|n,\xi,\zeta\rangle_{q^\dagger,q}$.

The amplitude we wish to compute is
\begin{eqnarray}
\langle B|n,\xi,\zeta\rangle_{\wt q,\wt q^\dagger}
&=&g_{\rm str}^{J+1}\int d^Ju
\Big\langle
\ cV_{\wt q}(-1)
\ V_Z(u_J)
\ \cdots
\nonumber\\&&
\quad\cdots
\ V_Z(u_1)
\ cV_{\wt q^\dagger}(1)
\ cV_{\rm excited}^{J(s=0)}(0)
\ \Big\rangle
+{\cal O}(g_{\rm str}^{J+2}).
\end{eqnarray}
If we use the normalization (\ref{unorm}) and
$\langle\partial Z^\dagger(u)Z(0)\rangle=1/u$,
this amplitude reduces to
\begin{eqnarray}
\langle B|n,\xi,\zeta\rangle_{\wt q,\wt q^\dagger}
&=&\frac{1}{4}
g_{\rm str}^{J+1}
O_0^{\wt q^\dagger}I_J
\oint\frac{du'}{2\pi i}\frac{1}{u{'}^n}
\Big\langle
S^{--}(-1)
S^{++}(+1)
{\cal F}_0(u',0)
\ \Big\rangle
\nonumber\\&&
+{\cal O}(g_{\rm str}^{J+2}).
\end{eqnarray}
The quantity ${\cal F}_0$, defined in (\ref{F0def}), consists of two terms,
a bosonic term and a fermionic term.
The bosonic term contribution is computed using the correlation function (\ref{bbonu}),
and we have
\begin{eqnarray}
&&\langle B|n,\xi,\zeta\rangle_{\wt q,\wt q^\dagger}^{\rm bosonic}
\nonumber\\
&&=\left\{\begin{array}{ll}
\displaystyle
\frac{1}{4}g_{\rm str}^{J+1}I_JO_0
(\xi_{\wt Z}\zeta_{\wt Z^\dagger}
+\xi_{\wt Z^\dagger}\zeta_{\wt Z}
-\xi_W\zeta_{W^\dagger}
-\xi_{W^\dagger}\zeta_W)
+{\cal O}(g_{\rm str}^{J+2}) & n=1,\\[2ex]
{\cal O}(g_{\rm str}^{J+2}) & n\geq2.
\end{array}\right.
\label{Aboson}
\end{eqnarray}
To compute the fermionic term contribution, we use the following correlation function of fermions in the $u$-plane with background $S^{--}(-1)$ and $S^{++}(1)$:
\begin{eqnarray}
\langle\psi^X(u_1)\psi^{X^\dagger}(u_2)\rangle
&=&\sqrt{\frac{1-u_1}{1+u_1}\frac{1+u_2}{1-u_2}}\left(\frac{1}{u_1-u_2}\mp\frac{1}{1-u_1u_2}\right),\nonumber\\
&&\left\{\begin{array}{l}
\mbox{$-$ for $X=Z$, $\wt Z$},\\
\mbox{$+$ for $X=W$}.
\end{array}\right.
\end{eqnarray}
Combining two propagators,
we obtain four fermion correlation functions as follows:
\begin{eqnarray}
&&\langle\psi^{\wt Z^\dagger}(u)\psi^{\wt Z}(0)\psi^{Z^\dagger}(u)\psi^Z(0)\rangle
  =\frac{1}{u^2}\frac{(1+u)^3}{1-u}\nonumber\\
&&\quad=\frac{1}{u^2}+\frac{4}{u}+7+8u+8u^2+8u^3+8u^4+\cdots+8u^n\cdots,\\
&&\langle\psi^{\wt Z}(u)\psi^{\wt Z^\dagger}(0)\psi^{Z^\dagger}(u)\psi^Z(0)\rangle
  =\frac{1}{u^2}(1-u^2),\\
&&\langle\psi^{W^\dagger}(u)\psi^W(0)\psi^{Z^\dagger}(u)\psi^Z(0)\rangle
  =\frac{1}{u^2}(1+u)^2,\\
&&\langle\psi^W(u)\psi^{W^\dagger}(0)\psi^{Z^\dagger}(u)\psi^Z(0)\rangle
  =\frac{1}{u^2}(1+u)^2.
\end{eqnarray}
From these, we have the fermionic term contribution
\begin{eqnarray}
&&\langle B|n,\xi,\zeta\rangle_{\wt q,\wt q^\dagger}^{\rm fermionic}
\nonumber\\
&&=\left\{\begin{array}{ll}
\displaystyle
\frac{1}{4}g_{\rm str}^{J+1}I_JO_0^{\wt q^\dagger}
(7\xi_{\wt Z^\dagger}\zeta_{\wt Z}
-\xi_{\wt Z}\zeta_{\wt Z^\dagger}
+\xi_{W^\dagger}\zeta_W
+\xi_W\zeta_{W^\dagger})
+{\cal O}(g_{\rm str}^{J+2}), & n=1,\\[2ex]
\displaystyle
2ng_{\rm str}^{J+1}I_JO^{\wt q^\dagger}_0\xi_{\wt Z^\dagger}\zeta_{\wt Z}
+{\cal O}(g_{\rm str}^{J+2}), & n\geq2.
\end{array}\right.
\label{Afermion}
\end{eqnarray}
The sum of (\ref{Afermion}) and (\ref{Aboson}) is
\begin{eqnarray}
\langle B|n,\xi,\zeta\rangle_{\wt q,\wt q^\dagger}
&=&\langle B|n,\xi,\zeta\rangle_{\wt q,\wt q^\dagger}^{\rm bosonic}
+\langle B|n,\xi,\zeta\rangle_{\wt q,\wt q^\dagger}^{\rm fermionic}
\nonumber\\
&=&2g_{\rm str}^{J+1}nI_JO^{\wt q^\dagger}_0\xi_{\wt Z^\dagger}\zeta_{\wt Z}
+{\cal O}(g_{\rm str}^{J+2}).
\end{eqnarray}

The other contribution of the spin $0$ term is obtained by
replacing the two twisting vertices $V_{\wt q}(-1)$ and $V_{\wt q^\dagger}(1)$
by $V_{q^\dagger}(-1)$ and $V_q(1)$, respectively,
and we have the following contribution proportional to
$O_{J+1}^{q^\dagger}$:
\begin{equation}
\langle B|n,\xi,\zeta\rangle_{q^\dagger,q}
=-2n(-)^ng_{\rm str}^{J+1}I_JO^{q^\dagger}_{J+1}\xi_{\wt Z^\dagger}\zeta_{\wt Z}
+{\cal O}(g_{\rm str}^{J+2}).
\end{equation}

Finally, we easily see that the spin $-1$ term
does not give a contribution of order $g_{\rm str}^{J+1}$.

Summing up all the contributions,
we obtain the following amplitude:
\begin{eqnarray}
\langle B|n,\xi,\zeta\rangle
&=&\langle B|n,\xi,\zeta\rangle_{s=1}
  +\langle B|n,\xi,\zeta\rangle_{\wt q,\wt q^\dagger}
  +\langle B|n,\xi,\zeta\rangle_{q^\dagger,q}\nonumber\\
&=&2g_{\rm str}^{J+1}nI_J\xi_{\wt Z}\zeta_{\wt Z}\quad\times\quad
                     \sum_{k=1}^JO^{\wt{\cal Z}}_k\cos(n\ol\theta_k)
\nonumber\\
      &+&2g_{\rm str}^{J+1}nI_J\xi_{\wt Z^\dagger}\zeta_{\wt Z}\quad\times\quad
                     \left(\sum_{k=1}^JO^{\wt{\cal Z}^\dagger}_k\cos(n\ol\theta_k)
                    +O^{\wt q^\dagger}_0
                    -(-)^nO^{q^\dagger}_{J+1}
                     \right)
\nonumber\\
     &-&2ig_{\rm str}^{J+1}nI_J\xi_W\zeta_{\wt Z}\quad\times\quad
                     \sum_{k=1}^JO^{\cal W}_k\sin(n\ol\theta_k)
\nonumber\\
     &-&2ig_{\rm str}^{J+1}nI_J\xi_{W^\dagger}\zeta_{\wt Z}\quad\times\quad
                     \sum_{k=1}^JO^{{\cal W}^\dagger}_k\sin(n\ol\theta_k)
\nonumber\\
&+&{\cal O}(g_{\rm str}^{J+2})+{\cal O}(J^{-1}),
\label{allcoupling}
\end{eqnarray}
where $\ol\theta_k=\pi k/(J+1)$.
This coupling precisely reproduces the correspondence (\ref{corresp2})--(\ref{corresp5}) obtained using gauge theory analysis,
up to the higher-order terms ${\cal O}(g_{\rm str}^{J+2})+{\cal O}(J^{-1})$.

To this point, we have considered the map from string states in the background without an
orientifold plane to operators in the $U(N_c)$ theory.
With this, it is straightfoward to obtain the map from string states in the orientifold to
gauge singlet operators in the $Sp(N_c/2)$ theory,
because the ${\bf Z}_2$ projections in gauge theory and string theory are
carried out consistently, as mentioned in \S\ref{oripro.ssec}.

\subsection{$1/J$ corrections}
In the computation of the disk amplitude given in \S\ref{excited.ssec},
we replaced the integrated $J$ coordinates $\theta_k$ in
(\ref{uintegral}) by
the even-interval points $\ol\theta_k=\pi k/(J+1)$.
This is possible when the angular momentum $J$ is sufficiently large.
In fact, the integral (\ref{inttk}) can be carried out analytically even
in case of finite $J$.
Therefore, it is interesting to compare the finite $J$ correction
in the gauge theory with that in the string theory.

The integral (\ref{inttk}) is carried out using the formula
\begin{equation}
\int_0^1\frac{t^{k-1}(1-t)^{J-k-1}}{(k-1)!(J-k-1)!}e^{zt}dt
=\frac{1}{(J-1)!}{}_1F_1(k,J,z),
\end{equation}
where ${}_1F_1(k,J,z)$ is the hypergeometric function defined by
\begin{equation}
{}_1F_1(k,J,z)=\sum_{m=0}^\infty
\frac{(k)_m}{(J)_m}\frac{z^m}{m!},\quad
(k)_m\equiv k\cdot(k+1)\cdots(k+m-1).
\end{equation}
When both $k$ and $J$ are sufficiently large,
we can replace $(k)_m$ and $(J)_m$ by $k^m$ and $J^m$, respectively,
and we obtain
\begin{equation}
{}_1F_1(k,J,z)
=\exp\frac{kz}{J}+{\cal O}(J^{-1}).
\end{equation}
If we ignore terms of order ${\cal O}(J^{-1})$,
we obtain the results given in \S\ref{excited.ssec}.
Inversely, we can obtain the exact result for finite $J$ by
replacing the triangular functions in (\ref{allcoupling})
by the hypergeometric functions according to
\begin{eqnarray}
\cos\left(\frac{\pi nk}{J+1}\right)\quad&\rightarrow&\quad
\frac{1}{2}[{}_1F_1(k,J+1,\pi in)+{}_1F_1(k,J+1,-\pi in)],
\label{hyper1}\\
\sin\left(\frac{\pi nk}{J+1}\right)\quad&\rightarrow&\quad
\frac{1}{2i}[{}_1F_1(k,J+1,\pi in)-{}_1F_1(k,J+1,-\pi in)].
\label{hyper2}
\end{eqnarray}
Unfortunately, the wave functions in the operators (\ref{corresp2})--(\ref{corresp5}) obtained in the gauge theory
and in (\ref{hyper1}) and (\ref{hyper2}) obtained in the string theory
do not coincide at finite $J$.

In Refs.\citen{ParSah} and \citen{ParRyz}, $1/J$ correction to the light-cone Hamiltonian of strings and
two-point functions of gauge singlet operators are studied,
and a result that is consistent in the string and gauge theories is obtained.
The ${\cal O}(1/J)$ correction we have obtained above is
different from the correction studied in Refs.\citen{ParSah} and \citen{ParRyz}.
They showed that mixing among operators with more than two insertions
can be reproduced in string theory by taking account of the finite radius correction to the PP-wave metric.
In our analysis, we consider operators with only one insertion and we do not use the curved background metric.
One possible origin of the disagreement between the $1/J$ corrections in the gauge theory and the string theory
is the mixing of states in the relation between $|n\rangle_{\rm circ}$ and $|n\rangle_{\rm AdS}$.
Although we cannot construct a string theory on a D3-brane solution background,
the semiclassical treatment used in Refs.\citen{GKP2,FroTse,ads} may be
useful in determining the corrections to the relation between
$|n\rangle_{\rm circ}$ and $|n\rangle_{\rm AdS}$.
We leave this problem to a future work.

\section{Conclusions}
In this paper we have investigated couplings between 7-7 strings and D3-branes.
We computed the inner products of boundary states of D3-branes
and the string circular wave states as disk amplitudes.
We showed that the ground state $b_{-1/2}^{\wt Z}|0\rangle$ and
the excited states $a_{-n}^ib_{-1/2}^{\wt Z}|0\rangle$ are mapped to the chiral operator
$\wt q{\cal Z}^{J-1}q$ and BMN operators in (\ref{corresp2})--(\ref{corresp5}),
respectively, by the relation (\ref{circ-on}).

The difference between the correspondence we obtain and that found in Ref.\citen{open}
is that our string states are circular wave states in flat spacetime
while those in Ref.\citen{open} are string states on the PP-wave background.
In order to obtain the correspondence between operators and string states in AdS,
we need to study the propagation of
strings on the D3-brane background.

We also discussed the $1/J$ correction to the operator-string state correspondence.
Unfortunately, the correction in the gauge theory and the result of the
string theory computation do not agree at
higher orders of $1/J$.

\section*{Acknowledgements}
The author would like to thank T.~Kawano and S.~Yamaguchi for variable 
discussions and useful comments.

%

\end{document}